\documentclass[envcountsame]{llncs}
\usepackage{amssymb}

\usepackage{times}

\title{There  exist some $\omega$-powers of any Borel rank} 
\author{Olivier Finkel\inst{1} \and Dominique Lecomte\inst{2}}
\institute{{\it Equipe Mod\`{e}les de Calcul et Complexit\'e}  
 \\ {\it Laboratoire de l'Informatique du Parall\'elisme}\footnote{UMR 5668 - CNRS - ENS Lyon - UCB Lyon - INRIA \\ LIP Research Report RR 2007-33}
 \\  CNRS et Ecole Normale Sup\'erieure de Lyon
 \\ 46, All\'ee d'Italie 69364 Lyon Cedex 07, France. \\ \email{Olivier.Finkel@ens-lyon.fr} \and 
{\it Equipe d'Analyse 
Fonctionnelle}
\\ Universit\' e Paris 6
\\4, place Jussieu, 75 252 Paris Cedex 05, France
 \\ \email{lecomte@moka.ccr.jussieu.fr}}

\date{}
\begin{document}

\spnewtheorem{Rem}[theorem]{Remark}{\bfseries}{\itshape}
\spnewtheorem{Exa}[theorem]{Example}{\bfseries}{\itshape}

\spnewtheorem{prop}[theorem]{Proposition}{\bfseries}{\itshape}
\spnewtheorem{lem}[theorem]{Lemma}{\bfseries}{\itshape}
\spnewtheorem{cor}[theorem]{Corollary}{\bfseries}{\itshape}
\spnewtheorem{defi}[theorem]{Definition}{\bfseries}{\itshape}

\def\ufootnote#1{\let\savedthfn\thefootnote\let\thefootnote\relax
\footnote{#1}\let\thefootnote\savedthfn\addtocounter{footnote}{-1}}

\newcommand{\bormxi}{{\bf\Pi}^{0}_{\xi}}
\newcommand{\bormlxi}{{\bf\Pi}^{0}_{<\xi}}
\newcommand{\bormz}{{\bf\Pi}^{0}_{0}}
\newcommand{\bormone}{{\bf\Pi}^{0}_{1}}
\newcommand{\ca}{{\bf\Pi}^{1}_{1}}
\newcommand{\bormtwo}{{\bf\Pi}^{0}_{2}}
\newcommand{\bormthree}{{\bf\Pi}^{0}_{3}}
\newcommand{\bormom}{{\bf\Pi}^{0}_{\omega}}
\newcommand{\borom}{{\bf\Delta}^{0}_{\omega}}
\newcommand{\borml}{{\bf\Pi}^{0}_{\lambda}}
\newcommand{\bormlpn}{{\bf\Pi}^{0}_{\lambda +n}}
\newcommand{\bormpm}{{\bf\Pi}^{0}_{1+m}}
\newcommand{\borapm}{{\bf\Sigma}^{0}_{1+m}}
\newcommand{\bormep}{{\bf\Pi}^{0}_{\eta +1}}
\newcommand{\borapxi}{{\bf\Sigma}^{0}_{\xi}}
\newcommand{\borai}{{\bf\Sigma}^{0}_{ 2.\xi +1 }}
\newcommand{\bormpxi}{{\bf\Pi}^{0}_{\xi}}
\newcommand{\bormpeta}{{\bf\Pi}^{0}_{1+\eta}}
\newcommand{\borapxipo}{{\bf\Sigma}^{0}_{\xi +1}}
\newcommand{\bormpxipo}{{\bf\Pi}^{0}_{\xi +1}}
\newcommand{\borpxi}{{\bf\Delta}^{0}_{\xi}}
\newcommand{\borel}{{\bf\Delta}^{1}_{1}}
\newcommand{\Borel}{{\it\Delta}^{1}_{1}}
\newcommand{\borone}{{\bf\Delta}^{0}_{1}}
\newcommand{\bortwo}{{\bf\Delta}^{0}_{2}}
\newcommand{\borthree}{{\bf\Delta}^{0}_{3}}
\newcommand{\boraone}{{\bf\Sigma}^{0}_{1}}
\newcommand{\boratwo}{{\bf\Sigma}^{0}_{2}}
\newcommand{\borathree}{{\bf\Sigma}^{0}_{3}}
\newcommand{\boraom}{{\bf\Sigma}^{0}_{\omega}}
\newcommand{\boraxi}{{\bf\Sigma}^{0}_{\xi}}
\newcommand{\ana}{{\bf\Sigma}^{1}_{1}}
\newcommand{\pca}{{\bf\Sigma}^{1}_{2}}
\newcommand{\Ana}{{\it\Sigma}^{1}_{1}}
\newcommand{\Boraone}{{\it\Sigma}^{0}_{1}}
\newcommand{\Borone}{{\it\Delta}^{0}_{1}}
\newcommand{\Bormone}{{\it\Pi}^{0}_{1}}
\newcommand{\Bormtwo}{{\it\Pi}^{0}_{2}}
\newcommand{\Ca}{{\it\Pi}^{1}_{1}}
\newcommand{\bormn}{{\bf\Pi}^{0}_{n}}
\newcommand{\bormm}{{\bf\Pi}^{0}_{m}}
\newcommand{\boralp}{{\bf\Sigma}^{0}_{\lambda +1}}
\newcommand{\borat}{{\bf\Sigma}^{0}_{|\theta |}}
\newcommand{\bormat}{{\bf\Pi}^{0}_{|\theta |}}
\newcommand{\Borapxi}{{\it\Sigma}^{0}_{\xi}}
\newcommand{\Bormpxipo}{{\it\Pi}^{0}_{1+\xi +1}}
\newcommand{\Borapn}{{\it\Sigma}^{0}_{1+n}}
\newcommand{\borapn}{{\bf\Sigma}^{0}_{1+n}}
\newcommand{\boraxipm}{{\bf\Sigma}^{0}_{\xi^\pm}}
\newcommand{\Boratwo}{{\it\Sigma}^{0}_{2}}
\newcommand{\Borathree}{{\it\Sigma}^{0}_{3}}
\newcommand{\Borapnpo}{{\it\Sigma}^{0}_{1+n+1}}
\newcommand{\Bormpxi}{{\it\Pi}^{0}_{\xi}}
\newcommand{\Borpxi}{{\it\Delta}^{0}_{\xi}}
\newcommand{\boratpxi}{{\bf\Sigma}^{0}_{2+\xi}}
\newcommand{\Boratpxi}{{\it\Sigma}^{0}_{2+\xi}}
\newcommand{\bormltpxi}{{\bf\Pi}^{0}_{<2+\xi}}
\newcommand{\Bormltpxi}{{\it\Pi}^{0}_{<2+\xi}}
\newcommand{\borapeap}{{\bf\Sigma}^{0}_{1+\eta_{\alpha ,p}}}
\newcommand{\borapeapn}{{\bf\Sigma}^{0}_{1+\eta_{\alpha ,p,n}}}
\newcommand{\Borapeap}{{\it\Sigma}^{0}_{1+\eta_{\alpha ,p}}}
\newcommand{\Bormpn}{{\it\Pi}^{0}_{1+n}}
\newcommand{\Borpn}{{\it\Delta}^{0}_{1+n}}
\newcommand{\borapximo}{{\bf\Sigma}^{0}_{1+(\xi -1)}}
\newcommand{\borpeta}{{\bf\Delta}^{0}_{1+\eta}}

\newcommand{\hs}{\hspace{12mm}

\noi}
\newcommand{\noi}{\noindent}
\newcommand{\ol}{ $\omega$-language}
\newcommand{\om}{\omega}
\newcommand{\Si}{\Sigma}
\newcommand{\Sis}{\Sigma^\star}
\newcommand{\Sio}{\Sigma^\omega}
\newcommand{\nl}{\newline}
\newcommand{\lra}{\leftrightarrow}
\newcommand{\fa}{\forall}
\newcommand{\ra}{\rightarrow}
\newcommand{\orl}{ $\omega$-regular language}

\maketitle

\begin{abstract}
\noi The operation $V \ra V^\om$  is a fundamental operation over finitary languages 
leading to \ol s. Since the set $\Sio$ of infinite words over a finite alphabet $\Si$ can be   equipped 
with the usual Cantor topology, the question of  the topological  complexity of  $\om$-powers of 
finitary  languages naturally arises and has  been posed by 
Niwinski \cite{Niwinski90}, Simonnet \cite{Simonnet92} and Staiger \cite{sta}. 
It has been recently  proved  that 
for each integer $n\geq 1$, there exist some $\om$-powers of  context free languages 
which are ${\bf \Pi}_n^0$-complete Borel sets,  \cite{Fin01a},   that there exists a 
context free language $L$ such that $L^\om$ is analytic but not Borel, \cite{Fin03a}, and 
 that  there exists a finitary language $V$ 
such that $V^\om$ is a Borel set of infinite rank,  \cite{Fin04}. But it was still  unknown which could be  the possible infinite Borel ranks of $\om$-powers. 
\nl 
We fill this gap here, proving the following very surprising result which shows that $\omega$-powers exhibit a great topological complexity: 
 for each non-null countable ordinal  $\xi$, there  exist 
some $\borapxi$-complete $\omega$-powers, and some  $\bormpxi$-complete  $\omega$-powers. 

\end{abstract}


\section{Introduction}

\noi
The operation $V \ra V^\om$  is a fundamental operation over finitary languages 
leading to \ol s. It produces $\omega$-powers, i.e.   \ol s in the form $V^\om$, where $V$ is a finitary language. 
This operation  appears  in the characterization of the class 
$REG_\om$  of  \orl s (respectively,  of the class $CF_\om$ of context free \ol s) 
 as the $\om$-Kleene closure 
of the family $REG$ of regular finitary languages (respectively,   of the 
family $CF$ of context free finitary languages) \cite{sta}.  
\nl Since the set $\Sio$ of infinite words over a finite alphabet $\Si$ can be   equipped 
with the usual Cantor topology, the question of  the topological  complexity of  $\om$-powers of 
finitary  languages naturally arises and has  been posed by 
Niwinski \cite{Niwinski90},  Simonnet \cite{Simonnet92},  and  Staiger \cite{sta}. 
A first task is to study the position of $\om$-powers 
with regard to the Borel hierarchy (and beyond to the projective hierarchy) \cite{sta,pp}. 

\hs It is easy to see that the  $\om$-power of a finitary language is always an analytic set because 
it is either the continuous image of a compact set $\{0,1, \ldots ,n\}^\om$ for $n\geq 0$
or of the Baire space $\om^\om$. 

\hs  It has been recently  proved,   that 
for each integer $n\geq 1$, there exist some $\om$-powers of  context free languages 
which are ${\bf \Pi}_n^0$-complete Borel sets,  \cite{Fin01a},  and   that there exists a 
context free language $L$ such that $L^\om$ is analytic but not Borel, \cite{Fin03a}. Notice that amazingly 
the language $L$ is  very simple to describe and it is accepted by a simple $1$-counter automaton. 

\hs  The first author proved in \cite{Fin04} that  there exists a finitary language $V$ 
such that $V^\om$ is a Borel set of infinite rank. However  the only known fact 
on their complexity is that there is a context free language $W$ such that $W^\om$ is Borel above 
${\bf \Delta_\omega^0}$, \cite{DF06}. 
\nl 
We fill this gap here, proving the following very surprising result which shows that $\omega$-powers exhibit a great topological complexity: 
 for each non-null countable ordinal  $\xi$, there  exist 
some $\borapxi$-complete $\omega$-powers, and some  $\bormpxi$-complete  $\omega$-powers. 
For that purpose we use a theorem of Kuratowski which is a level by level version of a theorem of Lusin and Souslin 
stating that every Borel set $B \subseteq 2^\om$ is the image of a closed subset of the Baire space 
$\om^\om$ by a continuous bijection. This theorem of Lusin and Souslin 
had already been used by Arnold in \cite{Arnold83} to prove that every Borel subset of 
$\Sio$, for a finite alphabet $\Si$,  is accepted by a non-ambiguous finitely branching  transition system with B\"uchi acceptance   
condition and our  first idea was to code the behaviour of such a transition system. This way, in the general case, 
we can manage to construct an $\om$-power of the same complexity as $B$. 

\hs The paper is organized as follows. In Section 2 we recall basic notions of topology and in particular definitions and properties of Borel sets.  
We proved our main result in Section 3.

\section{Topology}

\noi We first give some notations for finite or infinite words we shall use in the sequel, assuming  
the reader to be familiar with the theory of formal languages and of 
$\om$-languages, see \cite{tho,sta,pp}.  
\noi Let $\Si$ be a finite or countable alphabet whose elements are called letters.
A non-empty finite word over $\Si$ is a finite sequence of letters:
 $x=a_0.a_1.a_2\ldots a_n$ where $\fa i\in [0; n]$ $a_i \in\Si$.
 We shall denote $x(i)=a_i$ the $(i+1)^{th}$ letter of $x$
and $x\lceil (i+1)=x(0)\ldots x(i)$ for $i\leq n$, is the beginning of length $i+1$ of $x$. The length of $x$ is $|x|=n+1$.
The empty word will be denoted by $\emptyset$ and has 0 letters. Its length is 0.
 The set of finite words over $\Si$ is denoted $\Si^{<\om}$.
 A (finitary) language $L$ over $\Si$ is a subset of $\Si^{<\om}$.
 The usual concatenation product of $u$ and $v$ will be denoted by $u^\frown v$ or just  $uv$.
If $l\!\in\!\omega$ and $(a_{i})_{i<l}\!\in\! (\Si^{<\omega} )^l$, 
then ${^\frown}_{i<l}\ a_{i}$ is the concatenation $a_{0}\ldots a_{l-1}$.

\hs The first infinite ordinal is $\om$.
An $\om$-word over $\Si$ is an $\om$ -sequence $a_0a_1\ldots a_n \ldots$, where 
for all integers  $i\geq 0$~~ $a_i \in\Sigma$.  
 When $\sigma$ is an $\om$-word over $\Si$, we write
 $\sigma =\sigma(0)\sigma(1)\ldots  \sigma(n) \ldots $
and $\sigma \lceil (n+1)=\sigma(0)\sigma(1)\ldots  \sigma(n)$ the finite word of length $n+1$, 
prefix of $\sigma$.
The set of $\om$-words over  the alphabet $\Si$ is denoted by $\Si^\om$.
 An  $\om$-language over an alphabet $\Sigma$ is a subset of  $\Si^\om$.
If $\fa i\!\in\!\omega$ ~~ $a_{i}\!\in\! \Si^{<\omega} $, 
then ${^\frown}_{i\in\omega}\ a_{i}$ is the concatenation $a_0a_1\ldots$. 
 The concatenation product is also extended to the product of a 
finite word $u$ and an $\om$-word $v$: 
the infinite word $u.v$ or $u^\frown v$ is then the $\om$-word such that:
 $(u v)(k)=u(k)$  if $k < |u|$ , and  $(u.v)(k)=v(k-|u|)$  if $k\geq |u|$.
\nl 
 The prefix relation is denoted $\prec$: the finite word $u$ is a prefix of the finite 
word $v$ (respectively,  the infinite word $v$), denoted $u \prec v$,  
 if and only if there exists a finite word $w$ 
(respectively,  an infinite word $w$), such that $v=u^\frown  w$.
\nl If $s\!\prec\!\alpha\! =\!\alpha (0)\alpha (1)...$, then $\alpha\! -\! s$ is the sequence $\alpha (\vert s\vert )\alpha (\vert s\vert\! +\! 1)...$

\hs For a finitary language $V \subseteq \Si^{<\om}$, the $\om$-power of $V$ is the $\om$-language 
 $$V^\om = \{ u_1\ldots  u_n\ldots  \in \Si^\om \mid  \fa i\geq 1 ~~ u_i\in V \}$$

\noi We recall now some notions of topology, assuming  the reader to be familiar with basic notions  which
may be found in  \cite{ku,mos,kec,lt,sta,pp}.
\nl There is a natural metric on the set $\Sio$ of  infinite words 
over a countable alphabet 
$\Si$ which is called the prefix metric and defined as follows. For $u, v \in \Sio$ and 
$u\neq v$ let $d(u, v)=2^{-l_{pref(u,v)}}$ where $l_{pref(u,v)}$ is the first integer $n$
such that the $(n+1)^{th}$ letter of $u$ is different from the $(n+1)^{th}$ letter of $v$. 
The topology induced on  $\Sigma^\omega$ by this metric is just
 the product topology of the discrete topology on $\Sigma$. 
 For $s \in \Si^{<\om}$, the set 
$N_{s}\! :=\!\{\alpha\!\in\!\Sigma^\omega\mid s\!\prec\!\alpha\}$ 
is a basic clopen (i.e., closed and open) set of $\Sigma^\omega$. 
More generally open sets of $\Sio$ are in the form $W^\frown \Si^\om$, where $W\subseteq \Si^{<\om}$.

\hs The topological spaces in which we will work in this paper 
will be subspaces of $\Sigma^\omega$ where $\Si$ is either finite having at least two elements or countably infinite.
\nl When $\Si$ is a finite alphabet, 
the prefix  metric induces on $\Sio$ the usual  Cantor topology and $\Sio$ is compact. 
\nl The Baire space $\omega^\omega$ is equipped with the product topology of the 
discrete topology on $\omega$. It is homeomorphic to $P_\infty\! :=\!\{\alpha\!\in\! 2^\omega\mid
\forall i\!\in\!\omega\ \exists j\!\geq\! i\ \ \alpha (j)\! =\! 1\}\!\subseteq\! 2^\omega$, via the map 
defined on $\omega^\omega$ by $H(\beta )\! :=\! 0^{\beta (0)}10^{\beta (1)}1 \ldots$

\hs We define now the  {\bf Borel Hierarchy} on a  topological space $X$:

\begin{defi}
The classes ${\bf \Si}_n^0(X)$ and ${\bf \Pi}_n^0(X) $ of the Borel Hierarchy
 on the topological space  $X$ are defined as follows:
\nl ${\bf \Si}^0_1(X) $ is the class of open subsets of $X$.
\nl ${\bf \Pi}^0_1(X) $ is the class of closed subsets of $X$.
\nl And for any integer $n\geq 1$:
\nl ${\bf \Si}^0_{n+1}(X)$   is the class of countable unions 
of ${\bf \Pi}^0_n$-subsets of  $X$.
\nl ${\bf \Pi}^0_{n+1}(X)$ is the class of countable intersections of 
${\bf \Si}^0_n$-subsets of $X$.
\nl The Borel Hierarchy is also defined for transfinite levels.
The classes ${\bf \Si}^0_\xi(X)$
 and ${\bf \Pi}^0_\xi(X) $, for a non-null countable ordinal $\xi$, are defined in the
 following way:
\nl ${\bf \Si}^0_\xi (X)$ is the class of countable unions of subsets of $X$ in 
$\cup_{\gamma <\xi}{\bf \Pi}^0_\gamma $.
 \nl ${\bf \Pi}^0_\xi (X)$ is the class of countable intersections of subsets of $X$ in 
$\cup_{\gamma <\xi}{\bf \Si}^0_\gamma $.
\end{defi}

\noi  Suppose now that $X\!\subseteq\! Y$; then  
$\borapxi (X)\! =\!\{ A\cap X\mid A\!\in\!\borapxi (Y)\}$, and similarly for $\bormpxi$, see   \cite[Section 22.A]{kec}.
Notice that we have defined the Borel classes  ${\bf \Si}^0_\xi(X)$ and ${\bf \Pi}^0_\xi (X)$ mentioning the space $X$. 
However when the context is clear we will sometimes omit $X$ and denote ${\bf \Si}^0_\xi(X)$ by ${\bf \Si}^0_\xi$ and 
similarly for the dual class. 
\nl The Borel classes are closed under finite intersections and 
unions, and continuous preimages. Moreover, $\borapxi$ is closed under countable unions, and $\bormpxi$ 
under countable intersections. As usual  the ambiguous class  $\borpxi$ is the class $\borapxi\cap\bormpxi$.

\hs The class of 
{\bf Borel\ sets} is $\borel\! :=\!\bigcup_{\xi <\omega_1}\ \borapxi\! =\!
\bigcup_{\xi <\omega_1}\ \bormpxi$, where $\om_1$ is the first uncountable ordinal. 

\hs The {\bf Borel hierarchy} is as follows:
$$\begin{array}{ll}  
& \ \ \ \ \ \ \ \ \ \ \ \ \ \ \ \ \ \ \ \ \ \ \ \ \ \boraone\! =\!\hbox{\rm open}\ \ \ \ \ \ \ \ \ \ \ \ \ 
\boratwo\! \ \ \ \ \ \ \ \  \ \ \ 
\ldots\ \ \ \ \ \ \ \ \ \ \ \ \boraom\ \ \ \ \ \ldots\cr  
& \borone\! =\!\hbox{\rm clopen}\ \ \ \ \ \ \ \ \ \ \ \ \ \ \ \ \ \ \ \ \ \ \ \ \ \ \ 
\bortwo\ \ \ \ \ \ \ \ \ \ \ \ \ \ \ \ \ \ \ \ \ \ \ \ \ \ \ \ \ \ \ \borom\ \ \ \ \ \ \ \ \ \ \ \ \ \ \ \ \ \ \ \ \ \ \borel\cr
& \ \ \ \ \ \ \ \ \ \ \ \ \ \ \ \ \ \ \ \ \ \ \ \ \ \bormone\! =\!\hbox{\rm closed}\ \ \ \ \ \ \ \ \ \ \bormtwo\! \ \ \ \  \ \ \ \ \ \ \ \ \ldots
\ \ \ \ \ \ \ \ \ \ \ \ \bormom\ \ \ \ \ \ldots
\end{array}$$
This picture means that any class is contained in every class to the right of it, 
and the inclusion is strict in any of the spaces $\Sigma^\omega$. 

\hs For 
a countable ordinal $\alpha$,  a subset of $\Si^\om$ is a Borel set of {\it rank} $\alpha$ iff 
it is in ${\bf \Si}^0_{\alpha}\cup {\bf \Pi}^0_{\alpha}$ but not in 
$\bigcup_{\gamma <\alpha}({\bf \Si}^0_\gamma \cup {\bf \Pi}^0_\gamma)$.

\hs  We now define completeness with regard to reduction by continuous functions. 
For a countable ordinal  $\alpha\geq 1$, a set $F\subseteq \Si^\om$ is said to be 
a ${\bf \Si}^0_\alpha$  
(respectively,  ${\bf \Pi}^0_\alpha$)-{\it complete set} 
iff for any set $E\subseteq Y^\om$  (with $Y$ a finite alphabet): 
 $E\in {\bf \Si}^0_\alpha$ (respectively,  $E\in {\bf \Pi}^0_\alpha$) 
iff there exists a continuous function $f: Y^\om \ra \Si^\om$ such that $E = f^{-1}(F)$. 
  ${\bf \Si}^0_n$
 (respectively, ${\bf \Pi}^0_n$)-complete sets, with $n$ an integer $\geq 1$, 
 are thoroughly characterized in \cite{stac}.  
\nl Recall that 
a set $X \subseteq \Sio$ is a ${\bf \Si}^0_\alpha$
 (respectively ${\bf \Pi}^0_\alpha$)-complete subset of $\Sio$ iff it is in ${\bf \Si}^0_\alpha$ 
but not in ${\bf \Pi^0_\alpha}$  (respectively in  ${\bf \Pi}^0_\alpha$ but not in  ${\bf \Si}^0_\alpha$), \cite{kec}. 

\hs 
For example, the singletons of $2^\omega$ are $\bormone$-complete subsets of $2^\omega$. 
The set $P_\infty$ is a well known example of a $\bormtwo$-complete subset of $2^\omega$.

\hs  If ${\bf\Gamma}$ is a class of sets, then $\check {\bf\Gamma}\! :=\!\{\neg A\mid A\!\in\! {\bf\Gamma}\}$
 is the class 
of complements of sets in ${\bf\Gamma}$. In particular, for every 
 non-null countable ordinal $\alpha$, $\check {{\bf \Si}^0_\alpha}\!= {\bf \Pi}^0_\alpha$ and 
$\check {{\bf \Pi}^0_\alpha}\!= {\bf \Si}^0_\alpha$.  

\hs There are some subsets of the topological 
space $\Sio$   which are not Borel sets. In particular, there 
exists another hierarchy beyond the Borel hierarchy,  called the 
projective hierarchy. 
 The first class of the projective hierarchy is the class 
 ${\bf \Si}^1_1$ of {\bf analytic} sets. 
A set $A \subseteq \Sio$ is analytic 
iff there exists a Borel set $B \subseteq (\Si \times Y)^\om$, with $Y$  a finite alphabet,  
such that $ x \in A \lra \exists y \in Y^\om $ such that $(x, y) \in B$, 
where $(x, y)\in (\Si \times Y)^\om$ is defined by:  
$(x, y)(i)=(x(i),y(i))$ for all integers $i\geq 0$.  
\nl
A subset of $\Sigma^\omega$ is  analytic if it is empty, 
or the image of the Baire space by a continuous map. 
The class of analytic sets contains the class of Borel sets in any of the spaces $\Sigma^\omega$. 
Notice that ${\bf \Delta}_1^1 = {\bf \Si}^1_1 \cap {\bf \Pi}^1_1$, where ${\bf \Pi}^1_1$ is the class of co-analytic sets, i.e. of complements of 
analytic sets. 
 
\hs The $\om$-power of a finitary language $V$ is always an analytic set because if $V$ is finite and has $n$ elements then 
$V^\om$  is the continuous image of a compact set $\{0,1, \ldots ,n-1\}^\om$ and if $V$ is infinite then 
there is a bijection  between $V$ and $\om$ and $V^\om$  is the continuous image 
 of the Baire space $\om^\om$, 
\cite{Simonnet92}.

\section{Main result}

\noi  We now state our main result, showing that $\om$-powers exhibit a very surprising  topological complexity.

\begin{theorem}\label{main-result}

 Let $\xi$ be a non-null countable ordinal.\smallskip  

\noindent (a) There is $A\!\subseteq\! 2^{<\omega}$ such that $A^\om$ is $\borapxi$-complete.\smallskip

\noindent (b) There is $A\!\subseteq\! 2^{<\omega}$ such that $A^\om$ is 
$\bormpxi$-complete.

\end{theorem}
 
\noi To prove Theorem \ref{main-result},  we shall  use a level by level version of a theorem  of Lusin and Souslin 
stating that every Borel set $B \subseteq 2^\om$ is the image of a closed subset of the Baire space 
$\om^\om$ by a continuous bijection, see \cite[p.83]{kec}. 
It is the following theorem, proved by Kuratowski in \cite[Corollary 33.II.1]{ku}:
 
\begin {theorem}\label{kur}
 Let $\xi$ be a non-null countable ordinal, and $B\!\in\!\bormpxipo (2^\omega )$. 
Then there is $C\!\in\!\bormone (\omega^\omega )$ and a continuous bijection 
$f\! :\! C\!\rightarrow\! B$ such that $f^{-1}$ is $\borapxi$-measurable (i.e., $f[U]$ is $\borapxi (B)$ for each 
open subset $U$ of $C$).
\end{theorem} 

\noi The existence of the continuous bijection 
$f\! :\! C\!\rightarrow\! B$ 
given by this theorem (without the fact
that $f^{-1}$ is $\borapxi$-measurable) has been used by Arnold in \cite{Arnold83} to prove that every Borel subset of 
$\Sio$, for a finite alphabet $\Si$,  is accepted by a non-ambiguous finitely branching  transition system with B\"uchi acceptance   
condition. Notice that the sets of states of these transition systems are countable.
\nl Our first idea was to code the behaviour of such a transition system. In fact this can be done on a part of $\om$-words of a special 
compact set $K_{0,0}$. However  we shall have also to consider more general sets     $K_{N,j}$ and then we shall need the hypothesis of 
the $\borapxi$-measurability of the function $f$. 

\hs  We now  come to the proof of Theorem \ref{main-result}.

\hs   Let ${\bf\Gamma}$ be the class $\borapxi$, or $\bormpxi$. 
We assume first that $\xi\!\geq\! 3$.

\hs 
 Let $B \subseteq 2^\omega$ be a $ {\bf\Gamma}$-complete set. Then $B$ is in  $ {\bf\Gamma}(2^\omega)$ but not in   $\check {\bf\Gamma}(2^\omega)$. 
As $B\!\in\!\bormpxipo$, 
Theorem \ref{kur} gives $C \in\!\bormone (P_\infty)$ and $f$. By Proposition 11 in \cite{Lecomte05}, it is enough to find 
$A\!\subseteq\! 4^{<\omega}$. The language $A$ will be made of two pieces: we will 
have $A\! =\!\mu\cup\pi$. The set $\pi$ will code $f$, and $\pi^\om$ will look like $B$ on some nice compact sets  $K_{N,j}$. 
Outside this countable family of 
compact sets we will hide $f$, so that $A^\om$ will be the simple set $\mu^\om$. 

\hs  $\bullet$ We set $Q\! :=\!\{ (s,t)\!\in\! 2^{<\omega}\!\times\! 
2^{<\omega}\mid \vert s\vert\! =\!\vert t\vert\}$. We enumerate $Q$ as follows. We start 
with $q_{0}\! :=\! (\emptyset ,\emptyset )$. Then we put the sequences of length $1$ 
of elements of $2\!\times\! 2$, in the lexicographical ordering: 
$q_{1}\! :=\! (0,0)$, $q_{2}\! :=\! (0,1)$, $q_{3}\! :=\! (1,0)$, 
$q_{4}\! :=\! (1,1)$. Then we put the $16$ sequences of length $2$: 
$q_{5}\! :=\! (0^{2},0^{2})$, $q_{6}\! :=\! (0^{2},01)$, $\ldots$ And so on. We 
will sometimes use the coordinates of $q_{N}\! :=\! (q^{0}_{N},q^{1}_{N})$. We 
put $M_{j}\! :=\!\Sigma_{i<j}\ 4^{i+1}$. Note that the sequence 
$(M_j)_{j\in\omega}$ is strictly increasing, and that $q_{M_{j}}$ is the last 
sequence of length $j$  of elements of $2\!\times\! 2$.

\hs  $\bullet$ Now we define the ``nice compact sets". We will sometimes view $2$ as 
an alphabet, and sometimes view it as a letter. To make this distinction clear, we will use the boldface notation $\bf 2$ for the letter, 
and the lightface notation $2$ otherwise. 
We will have the same distinction with $3$ instead of $2$, so we have $2=\{0, 1\}, 3= \{0, 1, {\bf 2}\}, 
4=\{0, 1, {\bf 2}, {\bf 3}\}$. Let $N,j$ be non-negative integers with $N\!\leq\! M_{j}$. We set 
$$K_{N,j}:=\{\ \gamma = {\bf 2}^{N}\ {^\frown}\  [\ {^\frown}_{i\in\omega}\ \ m_{i}\ {\bf 2}^{M_{j+i+1}}\ {\bf 3}\ 
{\bf 2}^{M_{j+i+1}}\ ]\!\in\! 4^\omega  \mid  \fa i \in \om ~~m_i \in 2=\{0, 1\} \}.$$
As the map $\varphi_{N,j}\! :\! K_{N,j}\!\rightarrow\! 2^\omega$ defined by 
$\varphi_{N,j}(\gamma )\! :=\! {^\frown}_{i\in\omega} m_{i}$ is a homeomorphism, $K_{N,j}$ is 
compact.

\hs  $\bullet$ Now we will define the sets that ``look like $B$".

\hs  - Let $l\!\in\!\omega$. We define a function $c_l\! :\! B\!\rightarrow\! Q$ by 
$c_l(\alpha )\! :=\! [f^{-1}(\alpha ),\alpha ]\lceil l$. Note that $Q$ is countable, so that we equip it with the discrete topology. 
In these conditions, we prove that $c_l$ is $\borapxi$-measurable. 
\nl If $l\neq |q^0|=|q^1|$ then $c_l^{-1}(q)$ is the empty set. And for any $q\in Q$, and $l=|q^0|=|q^1|$, it holds that 
$c_l^{-1}(q)=\{ \alpha \in B \mid [f^{-1}(\alpha ),\alpha ]\lceil l = q\} =
 \{ \alpha \in B \mid \alpha \lceil l = q^1 \mbox{ and } f^{-1}(\alpha ) \lceil l = q^0 \}$. But  $\alpha \lceil l = q^1$ means that $\alpha$ belongs to the 
basic open set $N_{q^1}$ and $f^{-1}(\alpha ) \lceil l = q^0$ means that $f^{-1}(\alpha ) $ belongs to the 
basic open set $N_{q^0}$ or equivalently that $\alpha = f( f^{-1}(\alpha ) )$ belongs to $f(N_{q^0})$ which is a   $\borapxi$-subset of $B$. 
So  $c_l^{-1}(q) = N_{q^1}\cap f(N_{q^0})$ is a  $\borapxi$-subset of $B$ and $c_l$ is $\borapxi$-measurable. 

\hs  - Let $N$ be an integer. We put 
$$E_N\! :=\!\{\ \alpha\!\in\! 2^\omega\mid q^{1}_{N}\alpha\!\in\! B\ \ 
\hbox{\rm and}\ \ c_{\vert q^{1}_{N}\vert}(q^{1}_{N}\alpha )\! =\! q_{N}\ \}.$$ 
\noi Notice that $E_0=\{ \ \alpha\!\in\! 2^\omega\mid  \alpha\!\in B \mbox{ and } c_0(\alpha)=\emptyset \} = B$. 

\hs As $c_{\vert q^{1}_{N}\vert}$ is $\borapxi$-measurable and $\{ q_N\}\!\in\!\borone (Q)$, we get 
$c_{\vert q^{1}_{N}\vert}^{-1}(\{ q_N\})\!\in\!\borpxi (B)\!\subseteq\! {\bf\Gamma}(B)$. Therefore there is 
$G\!\in\!  {\bf\Gamma}(2^\omega )$ with $c_{\vert q^{1}_{N}\vert}^{-1}(\{ q_N\})\! =\! G\cap B$. Thus 
$c_{\vert q^{1}_{N}\vert}^{-1}(\{ q_N\})\!\in\! {\bf\Gamma}(2^\omega )$ since ${\bf\Gamma}$ is closed under finite intersections. 
Note that the map $S$ associating $q^{1}_{N}\alpha$ with $\alpha$ is continuous, so that 
$E_N\! =\! S^{-1}[c_{\vert q^{1}_{N}\vert}^{-1}(\{ q_N\})]$ is in 
${\bf\Gamma}(2^\omega )$.

\hs  $\bullet$ Now we define the transition system obtained from $f$.  

\hs - If $m\!\in\! 2$ and $n,p\!\in\!\omega$, then we write $n\buildrel m\over\rightarrow p$ if 
$q^{0}_{n}\!\prec\! q^{0}_{p}$ and $q^{1}_{p}\! =\! q^{1}_{n}m$.

\hs  - As $f$ is continuous on $C$, the graph $\hbox{\rm Gr}(f)$ of $f$ is a closed subset of 
$C\!\times\! 2^\omega$. As $C$ is $\bormone (P_\infty )$, $\hbox{\rm Gr}(f)$ is also a closed subset of $P_\infty\!\times\! 2^\omega$. 
So there is a closed subset $F$ of $2^\omega\!\times\! 2^\omega$ such that $\hbox{\rm Gr}(f)\! =\! F\cap (P_\infty\!\times\! 2^\omega )$. 
We identify 
$2^\omega\!\times\! 2^\omega$ with $(2\!\times\! 2)^\omega$, i.e., we view 
$(\beta ,\alpha )$ as $[\beta (0),\alpha (0)],[\beta (1),\alpha (1)],...$ By  \cite[Proposition 2.4]{kec}, there is 
$R\!\subseteq\! (2\!\times\! 2)^{<\omega}$, closed under initial segments, such that 
$F\! =\!\{ (\beta ,\alpha)\!\in\! 2^\omega\!\times\! 2^\omega\mid\forall k\!\in\!\omega\ \ 
(\beta ,\alpha )\lceil k\!\in\! R\}$; notice that $R$ is a tree whose infinite branches form the set $F$. 
In particular, we get 

$$(\beta ,\alpha )\!\in\!\hbox{\rm Gr}(f)\ \Leftrightarrow\ \beta\!\in\! P_\infty\ \ \hbox{\rm and}\ \ \forall k\!\in\!\omega\ \ (\beta,\alpha)\lceil k\!\in\! R.$$
- Set $Q_{f}\! :=\{ (t,s)\!\in\! R\mid t\!\not=\!\emptyset\ \ \hbox{\rm and}\ \ t(\vert t\vert\! -\! 1)\! =\!1\}$. Notice that $Q_f$ is simply the set of pairs 
$(t,s)\!\in\! R$ such that the last letter of $t$ is a $1$. 

\hs We have in fact already defined the transition system $\mathcal{T}$ obtained from $f$. 
This transition system has a countably  infinite set $Q$ of states and a 
set $Q_{f}$ of accepting states. The initial state is $q_{0}\! :=\! (\emptyset ,\emptyset )$. 
The input alphabet is $2=\{0, 1\}$ and the transition relation  $\delta \subseteq Q \times 2   \times Q$ is given 
by: if $m\!\in\! 2$ and $n,p\!\in\!\omega$ then $(q_n, m, q_p) \in \delta$ iff     $n\buildrel m\over\rightarrow p$.  
Recall that a run of $\mathcal{T}$ is said to be B\"uchi accepting if final states occur infinitely often during this run. 
Then the set of $\om$-words 
over the alphabet $2$ which are accepted by the transition system $\mathcal{T}$ from the initial state $q_0$ with B\"uchi acceptance condition 
is exactly the Borel set $B$. 

\hs  $\bullet$ Now we define the finitary language  $\pi$. 

\hs  - We set 
$$\pi\! :=\!\left\{ 
\begin{array}{ll}
& \!\!\!\!\!\! ~~~~s\!\in\! 4^{<\omega}\mid\exists j,l\!\in\!\omega\ \ 
\exists (m_i)_{i\leq l}\!\in\! 2^{l+1}\ \ 
\exists (n_i)_{i\leq l}, (p_i)_{i\leq l}, 
(r_i)_{i\leq l}\!\in\!\omega^{l+1}\cr 
& \cr
& \ \ \ \ \ \ \ \ \ \ \ \ \ \ \ \ \ \ \ \ \ \ \ \ \ \ \ \ \ \ \ \ \ \ \ 
\ \ \ \ \ \ \ \ \ \ \ \ \ \ \ \ \ \ \ \ \ \ \ \ \ \ \ \ \ \ \ \ \ \ 
\begin{array}{ll}
& \!\!\!\!\!\! n_{0}\!\leq\! M_j\cr & \hbox{\rm and}\cr
& \!\!\!\!\!\!\forall i\!\leq\! l\ \ n_i\buildrel 
{m_i}\over\rightarrow p_i\ \ \hbox{\rm and}\ \ 
p_i\! +\! r_i = M_{j+i+1}\cr & \hbox{\rm and}\cr
& \!\!\!\!\!\!\forall i\! <\! l\ \ p_i = n_{i+1}\cr & \hbox{\rm and}\cr
& \!\!\!\!\!\! q_{p_l}\!\in\! Q_f\cr & \hbox{\rm and}\cr
& \!\!\!\!\!\! 
s = {^\frown}_{i\leq l}\ \ {\bf 2}^{n_i}\ m_i\ {\bf 2}^{p_i}\ {\bf 2}^{r_i}\ {\bf 3}\ {\bf 2}^{r_i}
\end{array}\!\!\!
\end{array}\right\}.$$
$\bullet$ Let us show that $\varphi_{N,j}[\pi^\om\cap K_{N,j}]\! =\! E_N$ if 
$N\!\leq\! M_j$.\bigskip

 Let $\gamma\!\in\!\pi^\om\cap K_{N,j}$, and 
$\alpha\! :=\!\varphi_{N,j}(\gamma )$. We can write 

$$\gamma = {^\frown }_{k\in\omega}\ [\ {^\frown }_{i\leq l_{k}}\ \ {\bf 2}^{n^k_i}\ 
m^k_i\ {\bf 2}^{p^k_i}\ {\bf 2}^{r^k_i}\ {\bf 3}\ {\bf 2}^{r^k_i}\ ]\hbox{\rm .}$$

\noi As this decomposition of $\gamma$ is in $\pi$, we have 
$n^k_i\buildrel {m^k_i}\over\rightarrow p^k_i$ if $i\!\leq\! l_{k}$, 
$p^k_{i}\! =\! n^k_{i+1}$ if $i\! <\! l_{k}$, and $q_{p^k_{l_{k}}}\!\in\! Q_{f}$, 
for each $k\!\in\!\omega$. Moreover, $p^k_{l_{k}}\! =\! n^{k+1}_{0}$, for each 
$k\!\in\!\omega$, since $\gamma\!\in\! K_{N,j}$ implies that 
$p^k_{l_{k}}+r^k_{l_{k}}=r^k_{l_{k}}+n^{k+1}_{0}=M_{j+1+m}$ for some integer $m$. 
So we get 

$$N\buildrel {\alpha (0)}\over\rightarrow p^{0}_0
\buildrel {\alpha (1)}\over\rightarrow\ldots
\buildrel {\alpha (l_{0})}\over\rightarrow p^{0}_{l_{0}}
\buildrel {\alpha (l_{0}+1)}\over\rightarrow p^{1}_{0}
\buildrel {\alpha (l_{0}+2)}\over\rightarrow\ldots
\buildrel {\alpha (l_{0}+l_{1}+1)}\over\rightarrow p^{1}_{l_{1}}\ldots$$

\noi  In particular we have 

$$q^{0}_{N}\prec q^{0}_{p^{0}_0}\prec\ldots\prec q^{0}_{p^{0}_{l_{0}}}\prec 
q^{0}_{p^{1}_0}\prec\ldots\prec q^{0}_{p^{1}_{l_{1}}}\ldots$$

\noi  because $n \buildrel {m}\over\rightarrow p$ implies that $q_n^0 \prec q_p^0$. 
Note that $\vert q^{1}_{p^{k}_{l_{k}}}\vert\! =\!\vert q^{1}_{N}\vert\! +\!
\Sigma_{j\leq k}\ (l_{j}\! +\! 1)$ because  $n \buildrel {m}\over\rightarrow p$~  implies that $|q_p^1|=|q_n^1|+1$, so that the sequence 
$(\vert q^{0}_{p^{k}_{l_{k}}}\vert)_{k\in\omega}$ is strictly increasing since $|q_n^0|=|q_n^1|$ for each integer $n$. This implies 
the existence of $\beta\!\in\! P_{\infty}$ such that 
$q^{0}_{p^{k}_{l_{k}}}\prec\beta$ for each $k\!\in\!\omega$. Note that $\beta\!\in\! P_{\infty}$ because,  for each integer $k$,   
~$q_{p^{k}_{l_{k}}}\in Q_f$. Note also that 
$(\beta ,q^1_N\alpha )\lceil k\!\in\! R$ for infinitely many $k$'s. As $R$ is closed under initial segments, $(\beta ,q^1_N\alpha )\lceil k\!\in\! R$ for every $k\!\in\!\omega$, so that 
$q^1_N\alpha\! =\! f(\beta )\!\in\! B$. Moreover, 
$$c_{\vert q^{1}_{N}\vert}(q^{1}_{N}\alpha )\! =\! 
(\beta\lceil\vert q^{1}_{N}\vert ,q^{1}_{N})\! =\! 
(q^{0}_{N},q^{1}_{N})\! =\! q_{N}\hbox{\rm ,}$$ 
and $\alpha\!\in\! E_{N}$.\bigskip

\noi  Conversely, let $\alpha\!\in\! E_{N}$. We have to see that 
$\gamma\! :=\!\varphi_{N,j}^{-1}(\alpha )\!\in\!\pi^\om$. As $\gamma\!\in\! K_{N,j}$, 
we are allowed to write $\gamma = {\bf 2}^{N}\ {^\frown}\ [\ {^\frown}_{i\in\omega}\ \ \alpha (i)\ 
{\bf 2}^{M_{j+i+1}}\ {\bf 3}\ ^{M_{j+i+1}}\ ]$. Set $\beta\! :=\! 
f^{-1}(q^{1}_{N}\alpha )$. There is a sequence of integers $(k_{l})_{l\in\omega}$ such that 
${q_{k_{l}}\! =\! (\beta ,q^{1}_{N}\alpha )\lceil l}$. Note that 
$N\buildrel {\alpha (0)}\over\rightarrow k_{\vert q^{1}_{N}\vert +1}
\buildrel {\alpha (1)}\over\rightarrow k_{\vert q^{1}_{N}\vert +2}\ldots$ 
As $N\!\leq\! M_{j}$ we get $k_{\vert q^{1}_{N}\vert +i+1}\!\leq\! M_{j+i+1}$. So 
we can define $n_{0}\! :=\! N$, 
$p_{0}\! :=\! k_{\vert q^{1}_{N}\vert +1}$, 
$r_{0}\! :=\! M_{j+1}\! -\! p_{0}$, $n_{1}\! :=\! p_{0}$. Similarly, we can 
define $p_{1}\! :=\! k_{\vert q^{1}_{N}\vert +2}$, 
$r_{1}\! :=\! M_{j+2}\! -\! p_{1}$. We go on like this until we find some 
$q_{p_{i}}$ in $Q_{f}$. This clearly defines a word in $\pi$. And we can go 
on like this, so that $\gamma\!\in\!\pi^\om$.\bigskip

\noi  Thus $\pi^\om\cap K_{N,j}$ is in 
${\bf\Gamma}(K_{N,j})\!\subseteq\! {\bf\Gamma}(4^\omega )$. 
Notice that we proved, among other things, the equality $\varphi_{0,0}[\pi^\om\cap K_{0,0}]\! =\! B$. In 
particular, $\pi^\om\cap K_{0,0}$ is not in $\check {\bf\Gamma}(4^\omega )$.\bigskip

\noi Notice that $\pi^\om$ codes on $K_{0,0}$ the behaviour of the transition system accepting $B$. 
In a similar way     $\pi^\om$ codes on     $K_{N,j}$      the behaviour of the same  transition system     but starting this time from the state $q_N$ instead of the 
initial state $q_0$.                                    But some $\om$-words in $\pi^\om$ are not in 
$K_{0,0}$ and even not  in any $K_{N,j}$ and we do not know what is exactly the complexity of this set of $\om$-words. However we remark that all 
 words in $\pi$ have the same form 
${\bf 2}^N\ {^\frown}\ [\ {^\frown}_{i\leq l}\ \ m_i\ {\bf 2}^{P_i}\ {\bf 3}\ {\bf 2}^{R_i}\ ]$.

\hs  $\bullet$ We are ready to define $\mu$. The idea is that an infinite sequence 
containing a word in $\mu$ cannot be in the union of the $K_{N,j}$'s. We set 
$$\begin{array}{ll}
\mu^{0} & \!\!\!\! :=\!\left\{ 
\begin{array}{ll}
& \!\!\!\!\!\! ~~~~s\!\in\! 4^{<\omega}\mid\ \exists l\!\in\!\omega\ \ 
\exists (m_i)_{i\leq l+1}\!\in\! 2^{l+2}\ \ \exists N\!\in\!\omega
\ \ \exists (P_i)_{i\leq l+1}, (R_i)_{i\leq l+1}\!\in\!\omega^{l+2}\cr 
& \cr
& \ \ \ \ \ \ \ \ \ \ \ \ \ \ \ \ \ \ \ \ \ \ \ \ \ \ \ \ \ \ \ \ \ \ \ 
\ \ \ \ \ \ \ \ \ \ \ \ \ \ \ \ \ \ \ \ \ \ \ \ \ \ \ \ \ \ \ \ \ \ 
\begin{array}{ll}
& \!\!\!\!\!\!\forall i\!\leq\! l\! +\! 1\ \ \exists j\!\in\!\omega\ \  P_i\! =\!  M_{j}\cr & \hbox{\rm and}\cr
& \!\!\!\!\!\!  P_l\!\not=\! R_{l}\cr & \hbox{\rm and}\cr
& \!\!\!\!\!\! s = {\bf 2}^N\ {^\frown}\ [\ {^\frown}_{i\leq l+1}\ \ m_i\ {\bf 2}^{P_i}\ {\bf 3}\ {\bf 2}^{R_i}\ ]
\end{array}\!\!\!
\end{array}\right\}\hbox{\rm ,}\cr & \cr 
\mu^{1} & \!\!\!\! :=\!\left\{ 
\begin{array}{ll}
& \!\!\!\!\!\! ~~~~s\!\in\! 4^{<\omega}\mid\ \exists l\!\in\!\omega\ \ 
\exists (m_i)_{i\leq l+1}\!\in\! 2^{l+2}\ \ \exists N\!\in\!\omega
\ \ \exists (P_i)_{i\leq l+1}, (R_i)_{i\leq l+1}\!\in\!\omega^{l+2}\cr 
& \cr
& \ \ \ \ \ \ \ \ \ \ \ \ \ \ \ \ \ \ \ \ \ \ \ \ \ \ \ \ \ \ \ \ \ \ \ 
\ \ \ \ \ \ \ \ \ \ \ \ \ \ \ \ \ \ \ \ \ \ \ \ \ \ \ \ \ \ \ \ \ \ 
\begin{array}{ll}
& \!\!\!\!\!\!\forall i\!\leq\! l\! +\! 1\ \ \exists j\!\in\!\omega\ \  P_i\! =\!  M_{j}\cr & \hbox{\rm and}\cr
& \!\!\!\!\!\!\exists j\!\in\!\omega\ \  (P_l\!=\! M_{j}\ \ \hbox{\rm and}\ \ P_{l+1}\!\not=\! M_{j+1})\cr & \hbox{\rm and}\cr
& \!\!\!\!\!\! s = {\bf 2}^N\ {^\frown}\ [\ {^\frown}_{i\leq l+1}\ \ m_i\ {\bf 2}^{P_i}\ {\bf 3}\ {\bf 2}^{R_i}\ ]
\end{array}\!\!\!
\end{array}\right\}\hbox{\rm ,}\cr & \cr 
\mu & \!\!\!\! :=\!\mu^{0}\cup\mu^{1}.
\end{array}$$
All the words in $A$ will have the same form 
${\bf 2}^N\ {^\frown}\ [\ {^\frown}_{i\leq l}\ \ m_i\ {\bf 2}^{P_i}\ {\bf 3}\ {\bf 2}^{R_i}\ ]$. 
Note that any finite concatenation of words of this form still has this form. Moreover, 
such a concatenation is in $\mu^i$ if its last word is in $\mu^i$.\bigskip

\noindent $\bullet$ Now we show that $\mu^\om$ is ``simple". The previous remarks show that 

$$\mu^\om\! =\!\{\ \gamma\!\in\! 4^\omega\mid\exists i\!\in\! 2\ \ \forall j\!\in\!\omega\ \ 
\exists k, n\!\in\!\omega\ \ \exists t_0, t_1, \ldots, t_n\!\in\! \mu^i\ \ n \!\geq\! j
\ \ \hbox{\rm and}\ \ \gamma\lceil k\! =\! {^\frown}_{l \leq n}\ t_l\ \}.$$
This shows that $\mu^\om\!\in\!\bormtwo (4^\omega )$.

\hs Notice again that all 
 words in $A$ have the same form 
${\bf 2}^N\ {^\frown}\ [\ {^\frown}_{i\leq l}\ \ m_i\ {\bf 2}^{P_i}\ {\bf 3}\ {\bf 2}^{R_i}\ ]$. 
We set\bigskip 

\leftline{$P\! :=\!\{ {\bf 2}^N\ {^\frown}\ [\ {^\frown}_{i\in\omega}\ \ m_i\ {\bf 2}^{P_i}\ {\bf 3}\ {\bf 2}^{R_i}\ ]\!\in\! 4^\omega 
\mid  N\!\in\!\omega \mbox{ and } \fa i\in\omega ~~ m_i\!\in\! 2 , ~~  
P_i, R_i \in\omega  $}\medskip
\rightline{   and $ \forall i\!\in\!\omega\ \exists j\!\in\!\omega\ P_i\! =\! M_j\}.$}\bigskip

\noindent We define a map 
$F\! :\! P\!\setminus\!\mu^\om\!\rightarrow (\{\emptyset\}\cup\mu )\!\times\!\omega^2$ as follows. 
\nl 
Let $\gamma\! :=\! {\bf 2}^N\ {^\frown}\ 
[\ {^\frown}_{i\in\omega}\ \ m_i\ {\bf 2}^{P_i}\ {\bf 3}\ {\bf 2}^{R_i}\ ]\!\in\! P\!\setminus\!\mu^\om$, 
and $j_0\!\in\!\omega$ with $P_0\! =\! M_{j_0}$. If $\gamma\!\in\! K_{N,j_0-1}$,
 then we put $F(\gamma )\! :=\! (\emptyset ,N,j_0)$.  If $\gamma\!\notin\! K_{N,j_0-1}$, 
then there is an integer $l$ maximal for which $P_l\!\not=\! R_l$ or there is $j\!\in\!\omega$ with $P_l\! =\! M_j$ and 
$P_{l+1}\!\not=\! M_{j+1}$. Let $j_1\!\in\!\omega$ with $P_{l+2}\! =\! M_{j_1}$. We put
$$F(\gamma )\! :=\! ({\bf 2}^N\ {^\frown}\ [\ {^\frown}_{i\leq l}\ \ m_i\ {\bf 2}^{P_i}\ 
{\bf 3}\ {\bf 2}^{R_i}\ ]\ {^\frown}\ m_{l+1}\ {\bf 2}^{P_{l+1}}\ {\bf 3},R_{l+1},j_1).$$
$\bullet$ Fix $\gamma\!\in\! A^\om$. If $\gamma\!\notin\!\mu^\om$, 
then $\gamma\!\in\! P\!\setminus\!\mu^\om$, $F(\gamma )\! :=\! (t,S,j)$ is defined. Note that 
$t\ {\bf 2}^S\!\prec\!\gamma$, and that $j\! >\! 0$. Moreover, 
$\gamma\! -\! t\ {\bf 2}^S\!\in\! K_{0,j-1}$. Note also that 
$S\!\leq\! M_{j-1}$ if $t\! =\!\emptyset$, and that 
$t\ {\bf 2}^S\ \gamma (\vert t\vert\! +\! S)\ {\bf 2}^{M_{j}}\ {\bf 3}\!\notin\!\mu$. Moreover, there is an integer 
$N\!\leq\!\hbox{\rm min}(M_{j-1},S)$ ($N\! =\! S$ if $t\! =\!\emptyset$) such that 
$\gamma\! -\! t\ {\bf 2}^{S-N}\!\in\!\pi^\om\cap K_{N,j-1}$, 
since the last word in $\mu$ in the decomposition of $\gamma$ (if it exists) ends before $t\ {\bf 2}^S$.\bigskip 
 
\noindent $\bullet$ In the sequel we will say that $(t,S,j)\!\in\! (\{\emptyset\}\cup\mu )\times\omega^2$ is 
$\underline{suitable}$ if $S\!\leq\! M_{j}$ if $t\! =\!\emptyset$,  $t(\vert t\vert\! -\! 1)\! =\! {\bf 3}$ if 
$t\!\in\!\mu$, and $t\ {\bf 2}^S\ m\ {\bf 2}^{M_{j+1}}\ {\bf 3}\!\notin\!\mu$ if $m\!\in\! 2$. We set, for $(t,S,j)$ suitable, 
$$P_{t,S,j}:=\left\{\ \gamma\!\in\! 4^\omega\mid t\ {\bf 2}^S\!\prec\!\gamma\ \ \hbox{\rm and}\ \ 
\gamma\! -\! t\ {\bf 2}^S\!\in\! K_{0,j}\ \right\}.$$
Note that $P_{t,S,j}$ is a compact subset of $P\!\setminus\!\mu^\om$, and that 
$F(\gamma )\! =\! (t,S,j\! +\! 1)$ if $\gamma\!\in\! P_{t,S,j}$. This shows that the 
$P_{t,S,j}$'s, for $(t,S,j)$ suitable, are pairwise disjoint. Note also that $\mu^\om$ 
is disjoint from $\bigcup_{(t,S,j)\ \hbox{\rm suitable}}\ P_{t,S,j}$.\bigskip
   
\noindent $\bullet$ We set, for $(t,S,j)$ suitable and $N\!\leq\!\hbox{\rm min}(M_{j},S)$ 
($N\! =\! S$ if $t\! =\!\emptyset$),    
$$A_{t,S,j,N}:=\left\{\ \gamma\!\in\! P_{t,S,j}\mid\gamma\! -\! t\ {\bf 2}^{S-N}\!\in\!\pi^\om\cap K_{N,j}\ \right\}.$$ 
Note that $A_{t,S,j,N}\!\in\! {\bf\Gamma}(4^\omega )$ since $N\!\leq\! M_j$.\bigskip
  
\noindent $\bullet$ The previous discussion shows that 
$$A^\om\! =\!\mu^\om\cup
\bigcup_{(t,S,j)\ \hbox{\rm suitable}}
\bigcup_{
\begin{array}{ll}
& N\leq\hbox{\rm min}(M_j,S)\cr 
& \ N=S\ \hbox{\rm if}\ t=\emptyset
\end{array}}
\ A_{t,S,j,N}.$$ 

\noi As $\bf\Gamma$ is closed under finite unions, the set 

$$A_{t,S,j}:=\!\bigcup_{
\begin{array}{ll}
& N\leq\hbox{\rm min}(M_j,S)\cr 
& \ N=S\ \hbox{\rm if}\ t=\emptyset
\end{array}}
\ A_{t,S,j,N}$$

\noi 
is in ${\bf\Gamma}(4^\omega )$. 
On the other hand we have proved that  $\mu^\om \in\!\bormtwo (4^\omega )\!\subseteq\! {\bf\Gamma}(4^\omega )$, thus  we get 
$A^\om\!\in\! {\bf\Gamma}(4^\omega )$ if ${\bf\Gamma}\! =\!\borapxi$. 

\hs Consider now the case ${\bf\Gamma}\! =\!\bormpxi$. 
We can write 
$$A^\om\! =\!\mu^\om\!\setminus\!\left(\bigcup_{(t,S,j)\ \hbox{\rm suitable}}\ P_{t,S,j}
\right)\ \cup\bigcup_{(t,S,j)\ \hbox{\rm suitable}}\ A_{t,S,j}\cap P_{t,S,j}.$$

\noi 
Thus
$$\neg A^\om\! =\!\neg\left[ \mu^\om\cup\left(\bigcup_{(t,S,j)\ \hbox{\rm suitable}}\ P_{t,S,j}
\right)\right]\ \cup\bigcup_{(t,S,j)\ \hbox{\rm suitable}}\ P_{t,S,j}\!\setminus\! A_{t,S,j}.$$

\noi 
Here $\neg\left[ \mu^\om\cup\left(\bigcup_{(t,S,j)\ \hbox{\rm suitable}}\ P_{t,S,j}
\right)\right]\!\in\!\borthree (4^\omega )\!\subseteq\!\check {\bf\Gamma}(4^\omega )$ because $\mu^\om$ is a 
$\bormtwo$-subset of $4^\om$ and $(\bigcup_{(t,S,j)\ \hbox{\rm suitable}}\ P_{t,S,j})$ is a 
$\boratwo$-subset of $4^\om$ as it is a countable union of compact hence closed sets.  On the other hand  
$P_{t,S,j}\!\setminus\! A_{t,S,j}\!\in\!\check {\bf\Gamma}(4^\omega )$, thus $\neg A^\om$ is 
in $\check {\bf\Gamma}(4^\omega )$ and $A^\om\!\in\! {\bf\Gamma}(4^\omega )$. Moreover, the set 
$A^\om\cap P_{\emptyset ,0,0}\! =\!\pi^\om\cap P_{\emptyset ,0,0}\! =\!
\pi^\om\cap K_{0,0}$ is not in $\check {\bf\Gamma}$. This shows that $A^\om$ 
is not in $\check {\bf\Gamma}$. Thus $A^\om$ is in ${\bf\Gamma}(4^\omega )\!\setminus\!\check {\bf\Gamma}$.\bigskip

\noi We can now end the proof of Theorem \ref{main-result}. \bigskip

\noindent (a) If $\xi\! =\! 1$, then we can take 
$A\! :=\!\{ s\!\in\! 2^{<\omega}\mid 0\!\prec\! s\ \ \hbox{\rm or}\ \ 
\exists k\!\in\!\omega\ \ 10^k1\!\prec\! s\}$ and $A^\om\! =\! 2^\omega\!\setminus\!\{ 10^\om\}$ 
is $\boraone\!\setminus\!\bormone$.\bigskip

\noindent $\bullet$ If $\xi\! =\! 2$, then we will see in Theorem \ref{th2} the existence 
of $A\!\subseteq\! 2^{<\omega}$ such that $A^\om$ is $\boratwo\!\setminus\!\bormtwo$.\bigskip
 
\noindent $\bullet$ So we may assume that $\xi\!\geq\! 3$, and we are done.\bigskip 

\noindent (b) If $\xi\! =\! 1$, then we can take $A\! :=\!\{ 0\}$ and $A^\om\! =\!\{ 0^\om\}$ is 
$\bormone\!\setminus\!\boraone$.\bigskip

\noindent $\bullet$ If $\xi\! =\! 2$, then we can take $A\! :=\!\{ 0^k1\mid k\!\in\!\omega\}$ and $A^\om\! =\! P_\infty$ is $\bormtwo\!\setminus\!\boratwo$.\bigskip
 
\noindent $\bullet$ So we may assume that $\xi\!\geq\! 3$, and we are done.\hfill{$\square$}\bigskip

\noindent  As we have said above it remains a Borel class for which we have not yet got a complete $\om$-power:   the class $\boratwo$. 
Notice that it is easy to see that  the classical example of $\boratwo$-complete set, the set $2^\omega \setminus P_\infty$,  is not an $\om$-power. 
However we are going to prove the following result. 

\begin{theorem}\label{th2}
 There is a context-free  language $A\!\subseteq\! 2^{<\omega}$ such that 
$A^\om\!\in\!\boratwo\!\setminus\!\bormtwo$.
\end{theorem}

\noindent {\bf Proof.} By Proposition 11 in \cite{Lecomte05}, 
it is enough to find $A\!\subseteq\! 3^{<\omega}$. We set, for $j\! <\! 3$ and $s\!\in\! 3^{<\omega}$,
$$\begin{array}{ll}
n_j(s)\!\! & :=\ \hbox{\rm Card}\{ i\! <\!\vert s\vert\mid s(i)\! =\! j\}\hbox{\rm ,}\cr & \cr
\ \ \ \ \ \ T\!\! & :=\ \{\alpha\!\in\! 3^{\leq\omega}\mid\forall l\! <\! 1\! +\!\vert\alpha\vert\ \  
n_2(\alpha\lceil l)\!\leq\! n_1(\alpha\lceil l)\}.
\end{array}$$

\noi 
$\bullet$ We inductively define, for $s\!\in\! T\cap 3^{<\omega}$, $s^{\hookleftarrow}\!\in\! 2^{<\omega}$ as follows:
$$s^{\hookleftarrow}\! :=\!\left\{\!\!\!\!\!\!\begin{array}{ll}
&~~~~ \emptyset\ \ \hbox{\rm if}\ \ s\! =\!\emptyset\hbox{\rm ,}\cr & \cr
&~~~~ t^{\hookleftarrow}\varepsilon\ \ \hbox{\rm if}\ \ s\! =\! t\varepsilon\ \ \hbox{\rm and}\ \ 
\varepsilon\! <\! 2\hbox{\rm ,}\cr & \cr
&~~~~ t^{\hookleftarrow}\hbox{\rm ,\ except\ that\ its\ last\ 1\ is\ replaced\ with\ 0,\ if}\ s\! =\! t{\bf 2}.
\end{array}\right.$$

\noi 
$\bullet$ We will extend this definition to infinite sequences. To do this, we introduce a notion of limit. Fix $(s_n)_{n\in\omega}$ a sequence of elements in 
$ 2^{<\omega}$. We define 
${\displaystyle\lim_{n\rightarrow\infty}{s_n}}\!\in\! 2^{\leq\omega}$ as follows. For each 
$t\!\in\! 2^{<\omega}$,
$$t\!\prec\! {\displaystyle\lim_{n\rightarrow\infty}{s_n}}\ \Leftrightarrow\ \exists n_0\!\in\!\omega\ \ 
\forall n\!\geq\! n_0\ \ t\!\prec\! s_n.$$

\noi 
$\bullet$ If $\alpha\!\in\! T\cap 3^{\omega}$, then we set 
$\alpha^{\hookleftarrow}\! :=\! {\displaystyle\lim_{n\rightarrow\infty}{(\alpha\lceil n)^{\hookleftarrow}}}$. We define $e\! :\! T\cap 3^{\omega}\!\rightarrow\! 2^\omega$ by 
$e(\alpha )\! :=\!\alpha^{\hookleftarrow}$. Note that $T\cap 3^{\omega}\!\in\!\bormone (3^{\omega})$, and $e$ is a $\boratwo$-measurable partial function on $T\cap 3^{\omega}$, since for 
$t\!\in\! 2^{<\omega}$ we have 
$$t\!\prec\! e(\alpha )\ \Leftrightarrow\ \exists n_0\!\in\!\omega\ \ \forall n\!\geq\! n_0\ \ t\!\prec\! 
(\alpha\lceil n)^{\hookleftarrow}.$$

\noi 
$\bullet$ We set $E\! :=\!\{s\!\in\! T\cap 3^{<\omega}\mid n_2(s)\! =\! n_1(s)\ \ \hbox{\rm and}\ \ 
s\!\not=\!\emptyset\ \ \hbox{\rm and}\ \ 1\!\prec\! [s\lceil (\vert s\vert\! -\! 1)]^{\hookleftarrow}\}$. Note that 
$\emptyset\!\not=\! s^{\hookleftarrow}\!\prec\! 0^\om$, and that $s(\vert s\vert\! -\! 1)\! =\! {\bf 2}$ changes $s(0)\! =\! [s\lceil (\vert s\vert\! -\! 1)]^{\hookleftarrow}(0)\! =\! 1$ into $0$ if $s\!\in\! E$.\bigskip

\noindent $\bullet$ If $S\!\subseteq\! 3^{<\omega}$, then 
$S^*\! :=\!\{ {^\frown}_{i<l}\ s_i\!\in\! 3^{<\omega}\mid l\!\in\!\omega\ \ \hbox{\rm and}\ \ \fa  i<l ~~s_i \in  S\}$. We put 
$$A\! :=\!\{ 0\}\cup E\cup\{ {^\frown}_{j\leq k}\ (c_j1)\!\in\! 3^{<\omega}\mid
[\forall j\!\leq\! k\ \ c_j\!\in\! (\{ 0\}\cup E)^*]\ \ 
\hbox{\rm and}\ \ [k\! >\! 0\ \ \hbox{\rm or}\ \ (k\! =\! 0\ \ \hbox{\rm and}\ \ c_0\!\not=\!\emptyset )]\}.$$

\noindent $\bullet$ In the proof of Theorem \ref{main-result}.(b) we met 
the set $\{ s\!\in\! 2^{<\omega}\mid 0\!\prec\! s\ \ \hbox{\rm or}\ \ 
\exists k\!\in\!\omega\ \ 10^k1\!\prec\! s\}$. We shall denoted it by $B$ in the sequel. We have seen that  
 $B^\om\! =\! 2^\omega\!\setminus\!\{ 10^\om\}$ is 
$\boraone\!\setminus\!\bormone$. Let us show that $A^\om\! =\! e^{-1}(B^\om )$.\bigskip

\noindent - By induction on $\vert t\vert$, we get $(st)^{\hookleftarrow }={s^{\hookleftarrow }}
{t^{\hookleftarrow }}$ if $s,t\!\in\! T\cap 3^{<\omega}$. Let us show that 
$(s\beta)^{\hookleftarrow }\! =\! {s^{\hookleftarrow }}{\beta ^{\hookleftarrow }}$ if moreover 
$\beta\!\in\! T\cap 3^{\omega }$.\bigskip 

\noi  Assume that $t\!\prec\! (s\beta )^{\hookleftarrow }$. Then there is $m_{0}\!\geq\!\vert s\vert$ such that, for 
$m\geq m_{0}$, 
$$t\!\prec\! [(s\beta )\lceil m]^{\hookleftarrow }\! =\! [s\beta\lceil (m\! -\!\vert s\vert )]^{\hookleftarrow }\! =\! 
{s^{\hookleftarrow }}[\beta\lceil (m\! -\!\vert s\vert )]^{\hookleftarrow }.$$ 

\noi 
This implies that $t\prec {s^{\hookleftarrow }}{\beta ^{\hookleftarrow }}$ if 
$\vert t\vert\! <\!\vert s^{\hookleftarrow }\vert $. If $\vert t\vert\!\geq\!\vert s^{\hookleftarrow }\vert$, 
then there is $m_{1}\!\in\!\omega$ such that, for $m\!\geq\! m_{1}$, $\beta ^{\hookleftarrow }\lceil 
(\vert t\vert\! -\!\vert s^{\hookleftarrow }\vert )\!\prec\! [\beta\lceil (m\! -\!\vert s\vert )]^{\hookleftarrow }$. 
Here again, we get $t\!\prec\! {s^{\hookleftarrow }} {\beta ^{\hookleftarrow }}$. Thus 
$(s\beta)^{\hookleftarrow }\! =\! {s^{\hookleftarrow }}{\beta ^{\hookleftarrow }}$.\bigskip 

\noi  Let $(s_{i})_{i\in\omega}$ be a sequence such that for each integer $i\in\omega$, $s_i \in T\cap 3^{<\omega}$. 
Then ${^\frown}_{i\in\omega}\ s_{i}\!\in\! T$, and 
$({^\frown}_{i\in\omega}\ s_{i})^{\hookleftarrow }\! =\! {^\frown}_{i\in\omega}\ s_{i}^{\hookleftarrow }$, by the previous facts.\bigskip

\noindent - Let $(a_{i})_{i\in\omega}$ be a sequence such that for each integer $i\in\omega$, 
$a_{i}\in A\!\setminus\!\{\emptyset\}$ and 
$\alpha\! :=\! {^\frown}_{i\in\omega}\ a_i$. ~As $A\!\subseteq\! T$, 
$e(\alpha )\! =\! ({^\frown}_{i\in\omega}\ a_{i})^{\hookleftarrow }\! =\! {^\frown}_{i\in\omega}\ a_{i}^{\hookleftarrow }$. 
\nl If $a_0\!\in\!\{ 0\}\cup E$, then 
$\emptyset\!\not=\! a_0^{\hookleftarrow}\!\prec\!0^\om$, thus $e(\alpha )\!\in\! N_0\!\subseteq\! 2^\omega\!\setminus\!\{ 10^\om\}\! =\! B^\om$. 
\nl If $a_0\!\notin\!\{ 0\}\cup E$, then  
$a_0\! =\! {^\frown}_{j\leq k}\ (c_j1)$, thus 
$a_0^{\hookleftarrow}\! =\! {^\frown}_{j\leq k}\ (c_j^{\hookleftarrow}1)$. 

If $c_0\!\not=\!\emptyset$, then $e(\alpha )\!\in\! B^\om$ as before. 

If $c_0\! =\!\emptyset$, then $k\! >\! 0$, so that 
$e(\alpha )\!\not=\! 10^\om$ since $e(\alpha )$ has at least two coordinates

 equal to $1$. 
\nl We proved that $A^\om\!\subseteq\! e^{-1}(B^\om )$.\bigskip

\noindent - Assume now that $e(\alpha )\!\in\! B^\om$. We have to find 
$(a_i)_{i\in\omega}\!\subseteq\! A\!\setminus\!\{\emptyset\}$ with 
$\alpha\! =\! {^\frown}_{i\in\omega}\ a_i$. We split into cases:\bigskip

\noindent 1. $e(\alpha )\! =\! 0^\om$.

\noindent 1.1. $\alpha (0)\! =\! 0$.

\noi  In this case $\alpha\! -\! 0\!\in\! T$ and $e(\alpha\! -\! 0)\! =\! 0^\om$. Moreover, $0\!\in\! A$. We put 
$a_0\! :=\! 0$.\bigskip

\noindent 1.2. $\alpha (0)\! =\! 1$.

\noi  In this case there is a coordinate $j_0$ of $\alpha$ equal to ${\bf 2}$ ensuring that $\alpha (0)$ is replaced with a $0$ in $e(\alpha )$. We put $a_0\! :=\!\alpha\lceil (j_0\! +\! 1)$, so that  
 $a_0\!\in\! E\!\subseteq\! A$, $\alpha\! -\! a_0\!\in\! T$ and $e(\alpha\! -\! a_0)\! =\! 0^\om$.\bigskip
 
\noi  Now the iteration of the cases 1.1 and 1.2 shows that $\alpha\!\in\! A^\om$.\bigskip
 
\noindent 2. $e(\alpha )\! =\! 0^{k+1}10^\om$ for some $k\!\in\!\omega$.\bigskip

\noi  As in case 1, there is $c_0\!\in\! (\{ 0\}\cup E)^*$ such that 
$c_0\!\prec\!\alpha$, $c_0^{\hookleftarrow}\! =\! 0^{k+1}$, 
$\alpha\! -\! c_0\!\in\! T$ and $e(\alpha\! -\! c_0)\! =\! 10^\om$. Note that 
$\alpha (\vert c_0\vert )\! =\! 1$, $\alpha\! -\! (c_01)\!\in\! T$ and $e[\alpha\! -\! (c_01)]\! =\! 0^\om$. We put 
$a_0\! :=\! c_01$, and argue as in case 1.\bigskip
 
\noindent 3. $e(\alpha )\! =\! ({^\frown}_{j\leq l+1}\ 0^{k_j}1)0^\om$ for some $l\!\in\!\omega$.\bigskip

\noi  The previous cases show the existence of $(c_j)_{j\leq l+1}$, where for each $j\leq l+1$~ $c_j \in  (\{ 0\}\cup E)^*$ such that : 
\nl $a_0\! :=\! {^\frown}_{j\leq l+1}\ c_j1\!\prec\!\alpha$, $\alpha\! -\! a_0\!\in\! T$ and 
$e(\alpha\! -\! a_0)\! =\! 0^\om$. We are done since $a_0\!\in\! A$.\bigskip
 
\noindent 4. $e(\alpha )\! =\! {^\frown}_{j\in\omega}\ 0^{k_j}1$.\bigskip

\noi  An iteration of the discussion of case 3 shows that we can take $a_i$ of the form 
${^\frown}_{j\leq l+1}\ c_j1$.

\hs  $\bullet$ The previous discussion shows that $A^\om\! =\! e^{-1}(B^\om )$. 
As $B^\om$ is an open subset of $2^\om$ and 
$e$ is $\boratwo$-measurable, the $\om$-power $A^\om\! = e^{-1}(B^\om)$ is in $\!\boratwo (3^{\omega})$. 

\hs It remains to see that 
$A^\om\! = e^{-1}(B^\om)\!\notin\!\bormtwo$. We argue by contradiction.

\hs Assume on the contrary  that  $e^{-1}(B^\om)\!\in\!\bormtwo (3^{\omega})$. We know that  $B^\om\! =\! 2^\omega\!\setminus\!\{ 10^\om\}$ so 
$e^{-1}(\{10^\om\})= (T\cap 3^\om) \setminus  e^{-1}(B^\om)$ would be a $\boratwo$-subset of $ 3^\om$ since $T\cap 3^{\omega}$ is closed 
in $3^{\omega}$. 
Thus $e^{-1}(\{10^\om\})$ would be  a countable union of compact subsets of    $3^{\omega}$. 

\hs Consider now the {\bf cartesian product}  $(\{ 0\}\cup E)^{\mathbb{N}}$ of countably many copies of $(\{ 0\}\cup E)$. 
The set $(\{ 0\}\cup E)$ is countable and it can be equipped with the discrete 
topology. Then the product  $(\{ 0\}\cup E)^{\mathbb{N}}$ is equipped with the product topology 
of the discrete topology on $(\{ 0\}\cup E)$. The topological space 
$(\{ 0\}\cup E)^{\mathbb{N}}$ is homeomorphic to the Baire space $\om^\om$. 
\nl Consider now the map  $h\! :\! (\{ 0\}\cup E)^{\mathbb{N}}\!\rightarrow\! e^{-1}(\{10^\om\})$ defined by 
$h(\gamma )\! :=\! 1[{^\frown}_{i\in\omega}\ \gamma_i]$ for each 
$\gamma=(\gamma_0, \gamma_1, \ldots, \gamma_i, \ldots) \in (\{ 0\}\cup E)^{\mathbb{N}}$. 
The map $h$  is a homeomorphism by the previous discussion. As 
$(\{ 0\}\cup E)^{\mathbb{N}}$ is homeomorphic to the Baire space $\omega^\omega$, 
 the Baire space $\omega^\omega$ is also homeomorphic to  the space $e^{-1}(\{10^\om\})$,  so it would be also
a countable union of compact sets. 
But this is absurd by   \cite[Theorem 7.10]{kec}.

\hs It remains to see that $A$ is context-free. It is easy to see that the language $E$ is in fact accepted by a $1$-counter automaton: it is the set of words 
$s\!\in\!  3^{<\omega}$ such that :
$$\forall l \! \in [1;  \!\vert s \vert [ \ \  
n_2(s \lceil l)\!< \! n_1(s \lceil l) \mbox{ and }     n_2(s)\! =\! n_1(s)   \mbox{ and }   s(0)\! = 1 \mbox{ and }  s(\vert s\vert\! -\! 1)\! =\! {\bf 2}.$$ 
\noi This implies that $A$ is also accepted by a $1$-counter automaton because the class of $1$-counter languages is closed under concatenation and 
star operation. In particular $A$ is a context-free language because the class of languages accepted by  $1$-counter automata form a strict subclass 
of the class of context-free languages, \cite{ABB96}. 
\hfill{$\square$}

\begin{Rem}
The operation 
$\alpha \rightarrow \alpha^{\hookleftarrow}$ we have defined 
 is very close to the erasing operation defined by Duparc in his study of the Wadge hierarchy, \cite{Duparc01}. 
However we have  modified this operation in such a way that $\alpha^{\hookleftarrow}$ is always  infinite when $\alpha$ is infinite, and 
that it has the good property with regard to $\om$-powers and topological complexity.
\end{Rem}

\section{Concluding remarks and further work}

It is natural to wonder 
whether the $\om$-powers  obtained in this paper  are effective. For instance could they be obtained as $\om$-powers  of recursive languages ? 

\hs In the long version of this paper we prove effective versions of the results presented here. Using 
tools  of effective descriptive set theory, we first prove an effective version of Kuratowski's Theorem \ref{kur}. Then we use it 
to prove the following  effective version 
of Theorem \ref{main-result}, where $\Borapxi$ and $\Bormpxi$ denote classes of the hyperarithmetical hierarchy and $\omega_1^{CK}$ is the first 
non-recursive ordinal,  usually called the Church-kleene ordinal.

\begin{theorem} Let $\xi\!$ be a non-null ordinal smaller than $\omega_1^{CK}$.\smallskip 

\noindent (a) There is a recursive language $A\!\subseteq\! 2^{<\omega}$ such that 
$A^\om\!\in\!\Borapxi\!\setminus\!\bormpxi$.\smallskip

\noindent (b) There is a recursive language $A\!\subseteq\! 2^{<\omega}$ such that $A^\om\!\in\!\Bormpxi\!\setminus\!\borapxi$.\smallskip

\end{theorem}

\noi  The question, left open in \cite{Fin04},  also  naturally arises 
to know what are all the possible infinite Borel ranks of 
$\om$-powers of  finitary  
languages belonging to some natural class like the 
class of context free languages (respectively,   languages 
accepted by stack automata,  recursive languages, 
 recursively enumerable languages,  \ldots ). 
\nl We know from \cite{mscs06} that there are $\om$-languages accepted by B\"uchi $1$-counter automata of every Borel rank (and even of every 
Wadge degree) of an effective analytic set.  Every $\om$-language accepted by a B\"uchi $1$-counter automaton can be written as a finite union 
$L = \bigcup_{1\leq i\leq n} U_i^\frown V_i^\om$, where for each integer $i$, $U_i$ and $V_i$ are finitary languages accepted by $1$-counter automata. 
And the supremum of the set of Borel ranks of effective analytic sets is the ordinal 
 $\gamma_2^1$. 
This ordinal is defined by  A.S. Kechris, D. Marker, and  R.L. Sami in 
\cite{kms} and it is proved to be strictly greater than the ordinal 
$\delta_2^1$ which is the first non $\Delta_2^1$ ordinal. Thus the ordinal  $\gamma_2^1$ is also strictly greater than the first non-recursive 
ordinal  $ \om_1^{\mathrm{CK}}$.
From these results it seems  plausible that there exist some $\om$-powers of  languages accepted by  $1$-counter automata which have 
Borel ranks up to the ordinal  $\gamma_2^1$, although these languages are located at the very low level in the complexity hierarchy of finitary languages. 

\hs Another question concerns the Wadge hierarchy which is a great refinement of the Borel hierarchy. It would be interesting to determine the 
Wadge hierarchy of $\om$-powers. In the full version of this paper we give many  Wadge degrees of $\om$-powers and this confirms the 
great complexity of these $\om$-languages.

\end{document}